\documentstyle[12pt]{article}
\newcounter{ctr}
\begin{document}
\baselineskip=6mm
\begin{titlepage}
\begin{flushright}
   KOBE-TH-94-06\\
   Kanazawa-94-20\\
\end{flushright}
\vspace{2.3cm}
\centerline{{\large{\bf CP in Orbifold Models}}}
\par
\par
\par
\par\bigskip
\par\bigskip
\par\bigskip
\par\bigskip
\par\bigskip
\renewcommand{\thefootnote}{\fnsymbol{footnote}}
\centerline{{\bf Tatsuo Kobayashi}
\footnote[1]{e-mail:kobayasi@hep.s.kanazawa-u.ac.jp}
 and {\bf C.S. Lim}$^{\dagger}$
\footnote[3]{e-mail:lim@phys02.phys.kobe-u.ac.jp}}
\par
\par\bigskip
\par\bigskip
\centerline{Department of Physics, Kanazawa University, Kanazawa
920-11, Japan} \par
\centerline{Department of Physics, Kobe University, Nada, Kobe 657, Japan
$^{\dagger}$} \par
\par\bigskip
\par\bigskip
\par\bigskip
\par\bigskip
\par\bigskip
\par\bigskip
\par\bigskip
\par\bigskip
\par\bigskip
\centerline{{\bf Abstract}}\par

We study CP in orbifold models.
It is found that the orbifolds always have some automorphisms as CP
symmetry.
The symmetries are restricted non-trivially due to geometrical structure of
the orbifolds.
Explicit analysis on Yukawa couplings also shows that CP is not violated
in orbifold models.

\par\bigskip
\par\bigskip
\par\bigskip
\par\bigskip
\par\bigskip
\par\bigskip
\par\bigskip

\noindent September 1994
\end{titlepage}
\newpage

\vspace{0.8 cm}
\leftline{\large \bf 1. Introduction}
\vspace{0.8 cm}

To understand the origin of the CP violation is one of the most important
issues in the present particle physics. Though in the most popular and
successful Kobayashi-Maskawa model, for instance, CP violation is nicely
described by a single complex phase, we still seem to have not understood
deeply the reason why the phase appears and how the magnitude of it should
be as it really is.

The possibility of a new mechanism to break CP was pointed out to arise,
once the 4-dimensional standard model is regarded as the result
of dimensional reductions of higher dimensional theories, like
superstring or 10-dimensional supersymmetric Yang-Mills theories
\cite{MBGreen,CSLim}.
It was argued that whether CP symmetry in our 4-dimensional
space-time is broken or not crucially depends on the ways of
compactifications of the higher dimensional theories to the 4-dimensional
space-time. Thus, as the quite new feature, which never cannot be
expected in the ordinary 4-dimensional theories, the origin of CP symmetry
is reduced to the structure of the compactified space.

The new mechanism was explicitly confirmed to work in a
typical case of 10-dimensional superstring, investigating the
Yukawa couplings, and was shown to
be valid in other dimensionality, like 6-dimensional space-time, where
supersymmetric Yang-Mills theory can be defined as well \cite{CSLim}.
The importance
of the chiral nature of the theory, e.g. using Weyl (Majorana-Weyl)
fermion in the Type I superstring, in getting CP violation, was
also pointed out \cite{CSLim}.

In such chiral theories, naively defined P or both C and P in higher
dimensional space-time does not lead to the corresponding ones in
ordinary 4-dimensional space-time, and should be modified. The modified
operator, which corresponds to 4-dimensional CP operator, was demonstrated to
be equivalent to complex conjugation of the complex homogeneous
coordinates $z^a$ to describe the compactified space \cite{MBGreen},
\cite{CSLim}, i.e.
\begin{equation}
{\rm CP}: z^a \rightarrow z^{a*} .
\end{equation}
For instance in the four generation Calabi-Yau model with quintic
defining polynomial \cite{CY},
\begin{equation}
\sum_{a=1}^5 (z^a)^5 - C(z^1z^2 ... z^5) = 0,
\end{equation}
CP violation necessitates the complexity of the parameter $C$,
since otherwise
the defining equation will be invariant under the transformation
$z^a \rightarrow z^{a*}$. As for the 6-dimensional compactified space,
the complex conjugation of $z^a$ means that the corresponding
operator for the six real coordinates is orientation changing
\cite{MBGreen}, i.e. an operator whose matrix representation
has negative determinant: $diag (1,-1,1,-1,1,-1)$.

Recent discussions on the strong CP problem
show that CP symmetry based on the higher dimensional Yang-Mills
theories has a great advantage, when the effects of quantum gravity
are taken into account \cite{MDine}. In order to maintain the
smallness of the  $\theta$ parameter naturally, it seems to be
desirable to impose some sort of symmetry in weak-electromagnetic
sector. If the symmetry is global, however, the symmetry may be
spoilt once the quantum gravity effects are switched on. It
might be necessary to embed the symmetry in a local symmetry.
If we take CP itself as the symmetry, we may really regard it
as such a embedded symmetry, ``discrete gauge symmetry"
\cite{LMKrauss}, once the theory is embedded in a higher
dimensional Yang-Mills theories. For example, in 10-dimensional theory,
the CP operator has negative determinant, orientation changing, in
the compactified space.
\renewcommand{\thefootnote}{\fnsymbol{footnote}}
This means that, since the CP operator
in ordinary 4-dimensional space-time is also orientation changing, the
whole transformation in 10-dimensional space-time has positive
determinant, and can be embedded into a 10-dimensional Lorentz
transformation \cite{MDine}.\footnote[4]{Another solution to solve the
strong CP problem due to a Peccei-Quinn like symmetry in the superstring
theories is studied in refs.\cite{PQ}.}

Our purpose in the present paper is to discuss CP
in the superstring theories with $Z_N$ orbifold compactifications.
The orbifold construction is one of the simplest and most interesting
methods to derive 4-dimensional string vacua \cite{Orbi}.
In orbifold compactifications the compactified space is specified by
lattice vectors to define tori together with discrete symmetry groups to
devide
them, rather than the complex parameters like $C$ in Calabi-Yau
case. Since the CP operator is orientation changing, it causes
reflections in the subspaces of the compactified 6-dimensional space.
Thus, the problem of CP symmetry reduces to studying the symmetry
of the lattice vectors and their symmetry groups under the reflections.
If 6-dimensional orbifolds consist of three 2-dimensional orbifold,
transformation $z^a \rightarrow z^{a*}$ results in the reflections of each
2-dimensional subspace.
When we take a 6-dimensional orbifold which consists of three
2-dimensional $Z_2$ orbifolds, for example, the lattice obviously has the
reflection symmetry. A non-trivial complication arises, however, because
of the possible presence of non-trivial Wilson lines \cite{WL1,WL2,KO} and
structure of higher twisted sectors \cite{KO,KO2}.
We will discuss this point with special care.

We learn from the argument in Ref.2 that CP symmetry is completely
handled by the property of the compactified space (except for the
possibility of the spontaneous CP violation
in lower energies due to a complex VEV of
Higgs fields, which always cannot be ruled out). Thus, also in this
paper we first will consider whether the orbifolds have such
reflection symmetries or not for viable models to see whether
CP can be potentially violated, and will later
explicitly confirm our expectations
by use of concrete calculations of the Yukawa couplings of these
models. According to our results, CP is not broken in all models
under our consideration.

\vspace{0.8 cm}
\leftline{\large \bf 2. CP in higher dimensional theories}
\vspace{0.8 cm}

Before going into the detail of our discussion on the orbifold
models, we first briefly review the essence of the discussion on CP symmetry
in higher dimensional theories.
As has been pointed out in
our previous result \cite{CSLim}, the higher dimensional chirality
of the fermions plays an important role. In fact, in 10-dimensional
type \setcounter{ctr}{1}\Roman{ctr} superstring with
Majorana-Weyl fermion, CP behaves as $z^a \rightarrow z^{a*}$ as
was seen above for the compactified space, while in type
\setcounter{ctr}{2}\Roman{ctr}a theory with Majorana but
non-chiral fermions, CP can be defined naively just as ordinary one,
i.e. CP:$x^0 \rightarrow x^0,$ $x^\nu \rightarrow -x^\nu$
($\nu= 1 \sim 9$) for real coordinates, and CP is preserved even for the
complex
parameter C in the Calabi-Yau case, since the l.h.s of
the defining equation (2) is homogeneous quintic polynomial,
though we do not know any realistic model of this type.

The essential difference of these two cases comes from the fact
that the parity operator P in the compactified space acts as an
orientation changing one in type \setcounter{ctr}{1}\Roman{ctr}
case, while it acts as an orientation preserving one in type
\setcounter{ctr}{2}\Roman{ctr}a case. This difference can be
easily understood by the following simple argument. The 10-dimensional
fermions can be expressed in terms of $SO(10)$ momentum,
$(\pm \frac{1}{2},\pm \frac{1}{2},\pm \frac{1}{2},\pm \frac{1}{2},
\pm \frac{1}{2})$.
Since the 10-dimensional chirality of the fermion is determined
by $(-1)^n$, which we call as \lq \lq orientation" with $n$ being the
 number of entries $(-1/2)$, in type \setcounter{ctr}{1}\Roman{ctr}
model P has to preserve the 10-dimensional orientation.
On the other hand, P should change the 4-dimensional chirality, i.e.
the orientation of the first two entries of momentum.
Thus, the orientation of the remaining three entries for the
compactified space should be also changed.
In clear contrast to the above case, in type \setcounter{ctr}{2}
\Roman{ctr}a model, having both chiralities of 10-dimensional fermions,
the CP may change the chirality, and the naively defined CP operator,
which changes the 10-dimensional orientation while keeping the orientaion
of the compactified space, makes sense.

 In the realistic type \setcounter{ctr}{1}\Roman{ctr} theory, the
transformation in the compactified space discussed above, $z^a
\rightarrow z^{a*}$, is the typical example of the orientation
changing operator. We may, however, relax the definition of the
CP operator, as far as it is orientation changing \cite{MBGreen}.
For instance, a linear transformation of six real coordinates,
described by a matrix $diag (1,-1,1,1,1,1)$ has an equal right
to be a CP symmetry. This freedom is directly related with
the degree of freedom of a unitary transformation among four
4-dimensional spinors ($SU(4)$ internal symmetry) contained in a
10-dimensional Majorana-Weyl spinor.
We should note that the freedom of
Lorentz transformation $SO(6)$ is equivalent to $SU(4)$, and also
that 4-dimensional CP operator can be defined up to a freedom of a
unitary transformation. Thus the various orientation changing
operators in the 6-dimensional space, which are mutually
related via the freedom of $SO(6)$, have equal rights.
In practice, we may choose a suitable CP operator,
which is desirable to discuss the each case of compactification,
like $z_a \rightarrow z_a^*$ in the Calabi-Yau case, and as
far as we can find even one CP operator under which the
compactification is invariant, 4-dimensional CP
will not be physically broken.

\vspace{0.8 cm}
\leftline{\large \bf 3. CP in orbifold models}
\vspace{0.8 cm}

In the orbifold models, a state consists of a bosonic string on the
4-dimensional space-time and the 6-dimensional orbifold $x^\mu$
($\mu=0 \sim 9$), their right-moving superpartner $\psi^\mu$ and a
left-moving gauge part $x^I$ $(I=1 \sim 16)$.
The orbifold is a division of a torus by a twist $\theta$.
We denote $\theta$ in a complex basis $z^a$ ($a=2 \sim 4$) as
$\exp[2 \pi i v^a]$, where $z^t=x^{2t}+ix^{2t+1}$.
Here we restrict ourselves to the 6-dimensional $Z_N$ orbifolds which are
products of 2-dimensional orbifolds, i.e., $Z_N$ orbifolds $(N=3,4,6)$.

A $\theta^k$-twisted string has the following boundary condition,
$$x^i(\sigma=2\pi)=\theta^kx^i(\sigma=0)+e^i,
\eqno(3)$$
where $e^i$ is a lattice vector to define the torus.
A zero mode of the twisted string satisfy the same equation as (3) and
is called a fixed point.
The fixed point is denoted by the space group element
$(\theta^k,e^i)$.
It is remarkable that all fixed points in the higher twisted sectors are not
fixed under $\theta$.
\renewcommand{\thefootnote}{\fnsymbol{footnote}}
For example the 2-dimensional $Z_6$ orbifold has fixed points
$(\theta^2,e_1)$,
$(\theta^2,e_1+e_2)$, $(\theta^3,e_1)$, $(\theta^3,e_1+e_2)$ and
$(\theta^3,e_2)$ as well as the origin \footnote[5]{See ref.\cite{KO} for
several geometrical aspects of the orbifolds.},
where the vectors $e_1$ and $e_2$
represent two simple roots of $SU(3)$ and $\theta=\exp [\pi i/3]$.
The first two fixed points are transformed each other by $\theta$.
The others are also transformed by the $\theta$ twist as,
$$ (\theta^3,e_1) \rightarrow (\theta^3,e_1+e_2) \rightarrow
(\theta^3,e_2) \rightarrow (\theta^3,e_1).
\eqno(4)$$
In this case we have to take linear combinations of the corresponding
states to construct eigenstates of $\theta$ \cite{KO2,WL2,KO}.
For the $\theta^3$-twisted sector of the $Z_6$ orbifold we have the following
eigenstates of $\theta$ up to a normalization factor,
$$|\theta^3,e^{i \gamma}>=|(\theta^3,e_1)>+e^{2i\gamma}
|(\theta^3,e_1+e_2)>+e^{i\gamma}|(\theta,e_2)>,
\eqno(5)$$
where $e^{i\gamma}$ should be one of the third roots of unity.
This state have eigenvalue of the $Z_N$ twist as $e^{i\gamma}$.
Similarly the 2-dimensional $Z_4$ orbifold has linearly combined states.

We can bosonize $\psi^\mu$.
In the $\theta^k$-twisted sector, the bosonized field has a momentum as
$\tilde p^t=p^t+kv^t$ ($t=0 \sim 4$), where $p^t$ is located on the $SO(10)$
weight lattice and $v^t=0$ for $t=0,1$, corresponding to the 4-dimensional
space-time.
Similarly the left-moving gauge part of the $\theta^k$-twisted sector has
a momentum $\tilde P^I=P^I+kV^I+a_{e^i}^I$ ($I=1 \sim 16$), where $P^I$ is
located on an $E_8\times E_8$ lattice, $V^I$ is a shift vector and
$a_{e^i}^I$ is a Wilson line corresponding to $e^i$ \cite{WL1,WL2,KO}.

A massless state associated with $(\theta^k,e^i)$ should satisfy the
following conditions in the light-cone gauge,
$${1 \over 2}\sum_{t=1}^4(\tilde p^t)^2+N_k+c_k={1\over 2}, \quad
{1 \over 2}\sum_{I=1}^{16}(\tilde P^I)^2+N_k+c_k=1,
\eqno(6)$$
where $N_k$ is an oscillator number and $c_k$ is a ground state energy of
the $\theta^k$ twisted sector.
Further physical states are invariant under a total $Z_N$ transformation.
A $Z_N$ phase $\Delta$ under the transformation is obtained as \cite{WL2,KO}
$$
\begin{array}{cl}
\Delta=& O^{(k)}e^{i\gamma}\exp 2 \pi i[ {k \over 2} \{ \displaystyle
\sum_t(v^t)^2
-\sum_I(V^I+a_{e^i}^I)^2 \} \cr
& +\displaystyle
\sum_I(V^I+a_{e^i}^I)(P^I+kV^I+a_{e^i}^I) -\sum_t(p^t+kv^t)v^t],
\end{array}
\eqno(7)
$$
where $O^{(k)}$ is a contribution of oscillated states.
Here $e^{i\gamma}$ is the phase obtained, when we take linear combinations
of the 6-dimensional ground states associated with the fixed points.
Physical states should have $\Delta=1$.
Hereafter a phase $\tilde \Delta$ represents
$e^{-i \gamma} \Delta /O^{(k)}$.

Now we study CP on the orbifold models.
The parity operator P should be some transformation of the 6-dimensional
space whose determinant is --1 and C is the charge conjugation, i.e.,
C:$\tilde P^I \rightarrow -\tilde P^I$ \cite{MBGreen,CSLim,MDine}.
The former corresponds to some kinds of \lq reflections', e.g.,
$z^a \rightarrow z^{a*}$ or $z^a \leftrightarrow z^b$ ($a \neq b$).
The CP can be defined well if both of the massless conditions and the
$Z_N$ phase $\Delta$ are invariant under the CP.
Before compactifications, several kinds of reflections are candidates for
P.
However, compactifications forbid most of them.
Here we consider P where all $z^a$ are transformed into $z^{a*}$ as well as
the 4-dimensional parity reflection.
Other possibilities will be discussed later.
This transformation on $z^a$ induces a transformation of the $SO(10)$ momenta
as
$$\tilde p^t \rightarrow -(p^0,-\tilde p^1,\tilde p^2,\tilde p^3,
\tilde p^4)=-(p^0,-p^1,p^2+kv^2,p^3+kv^3,p^4+kv^4),
\eqno(8)$$
because $x^\mu$ is related with $\psi^\mu$ through the
world-sheet supersymmetry.
Eq.(8) implies that P transforms the $\theta^k$-twisted sector into the
$\theta^{N-k}$-twisted sector in the $Z_N$ orbifold model.
At the same time $\tilde P^I$ should be transformed into
$-\tilde P^I=-P^I-kV^I-a_{e}^I$, i.e., the charge conjugation C.
These transformations are also justified from the fact that there is a term
$g_{\mu I}\partial x^\mu \partial x^I$ in the lagrangian when the Wilson
lines exist.
The C and P transformations should be taken simultaneously so that
the term in the lagrangian is not changed.
Thus the state associated with $(\theta^k,e)$ is transformed
under the CP transformation as follows,
$${\rm CP}:|(\theta^k,e)> \rightarrow |(\theta^{N-k},-e)>.
\eqno(9)$$
The massless conditions are invariant under the transformation.
In addition the phase $\tilde \Delta$ is transformed into $1/\tilde \Delta$.
That requires that $e^{i \gamma}$ should be transformed into
$e^{-i \gamma}$ in order to lead to $\Delta=1$.

Here we study geometrically reflections of the orbifolds which transform
the fixed point $(\theta^k,e)$ into $(\theta^{N-k},-e)$ up to the
conjugacy class.
For example we consider fixed points on the 2-dimensional $Z_3$ orbifold,
whose lattice are spanned by two simple roots of $SU(3)$, i.e.,
$e_1=(\sqrt 2,0)$ and $e_2=(-1/\sqrt 2,\sqrt {3/2})$.
There are two fixed points $(\theta,e_1)$ and $(\theta,e_1+e_2)$ as well as
the origin, where $\theta=\exp [2 \pi i/3]$.
In the $Z_3$ orbifold the space group element $(\theta,e_2)$ is
equivalent to $(\theta,e_1)$ and we obtain $3e_1 \approx 0$ up to $\theta$.
This $SU(3)$ lattice has some types of automorphisms, e.g., a reflection
of each element of $e_1$ and $e_2$.
A reflection, which changes the sign of the first coordinates of $e_1$
and $e_2$, transforms the simple roots as follows,
$$e_1 \rightarrow -e_1,\quad e_1+e_2\ (\approx 2e_1) \rightarrow e_2 \
(\approx -2e_1).
\eqno(10)$$
Thus it works as a well defined P.
On the other hand, a reflection, which changes the sign of the second
coordinates, transforms $e_1$ and
$e_1+e_2$ into $e_1$ and $-e_2$, respectively and it does not work as P.
The former corresponds to a Weyl reflection of $SU(3)$.
We need the following remark.
If Wilson lines vanish, we can not distinguish $(\theta,e_1)$ and
$(\theta,e_1+e_2)$.
Thus in this case any automorphism including the non-Weyl reflection
can be defined as P.

The 2-dimensional $Z_4$ and $Z_6$ orbifolds are constructed through the
$SO(4)$ and $SU(3)$ lattices, respectively.
The $Z_6$ orbifold has no fixed point except the origin under
$\theta$-twist, where $\theta=\exp [\pi i /3]$.
Therefore any automorphism of the lattice works as a well defined P.
Similarly any automorphism of the $SO(4)$ lattice can be used as P, because
the $\theta$-twisted sector of the $Z_4$ orbifold has only one fixed point
$(\theta,e_1+e_2)$ as well as the origin, where $\theta=\exp [\pi i/2]$ and
$e_1$ and $e_2$ are ($1,0$) and ($0,1$), respectively.
It is remarkable that the Wilson lines are not allowed in the 2-dimensional
$Z_6$ orbifold and the 2-dimensional $Z_4$ orbifold is possible to have the
Wilson lines with order two \cite{WL2,KO}.

Next we discuss the 2-dimensional $Z_2$ orbifold whose lattice is spanned by
$e_1$ and $e_2$.
The orbifold has three fixed points $(\theta,e_1)$, $(\theta,e_1+e_2)$ and
$(\theta,e_2)$ as well as the origin, where $\theta=\exp [\pi i]$.
A reflection along $e_1$ or $e_2$ works for P.
If $e_1$ corresponds to the same Wilson line as $e_2$, an exchange between
$e_1$ and $e_2$ can also be used for P.

Both the $\theta^2$-twisted sector of the $Z_4$ orbifold and
the $\theta^3$-twisted sector of the $Z_6$ orbifold have the same structure
of fixed points as the $Z_2$ orbifold.
However, we have to consider more detail, because these fixed points are
not fixed under $\theta$ as said before.
In this case we have to take linear combinations of states associated with
fixed points like in eq.(5).
The operator P should transform $|\theta^3,e^{i \gamma}>$ as follows,
$$|\theta^3,e^{i \gamma}> \rightarrow |\theta^3,e^{-i \gamma}>,
\eqno(11)$$
in order to be consistent with the GSO projection.
Therefore P should transform $(\theta,e_1)$ into $(\theta,e_2)$ and vice
versa, while $(\theta,e_1+e_2)$ is not changed.
\renewcommand{\thefootnote}{\fnsymbol{footnote}}
This is nothing but a Weyl reflection of $G_2$.\footnote[6]
{The 2-dimensional orbifold $Z_6$ is defined by a $G_2$ lattice and its
Coxeter element as a twist $\theta$ \cite{KO}.}
Thus the candidates of P are restricted severely by considering the higher
twisted sectors, although the $\theta$-twisted sector allows some types
of P.
Next we have to study the $\theta^2$-twisted sector of the $Z_6$
orbifold.
The above reflection is also consistent with the $\theta^2$-twisted sector.
Similarly we discuss the case of the $\theta^2$-twisted sector of the $Z_4$
orbifold.
The operator P should be a Weyl reflection of SO(4).

We have studied the 2-dimensional orbifolds.
The 6-dimensional orbifolds are obtained by taking direct products of
three 2-dimensional orbifolds.
For example the $Z_6$-II orbifold consists of the 2-dimensional $Z_2$, $Z_3$
and $Z_6$ orbifolds.
For these 6-dimensional orbifolds, the well defined P's are products of the
Weyl reflections discussed above.

We have considered the case where P transformation $z^a \rightarrow z^{a*}$
causes simultaneous Weyl reflections in all 2-dimensional orbifolds.
In general only the simultaneous reflections of the 2-dimensional spaces are
consistent with (6) and (7).
However, some cases allow partial reflections.
For example, we discuss the case where the 6-dimensional orbifolds like $Z_4$
and $Z_6$ include the 2-dimensional $Z_2$ orbifolds.
In this case, a well difined CP is such that the Weyl reflection is only
for the 2-dimensional $Z_2$ space, while the other 4-dimensional space is
fixed.
This is related with the fact that these orbifolds have partially $N=2$
supersymmetric sectors.
Some 6-dimensional orbifolds have the same eigenvalues of the twists,
e.g., $v^2=v^3$.
In this case, an exchange between $z^2$ and $z^3$ can also work as P.

The above study shows that CP is a good symmetry in the orbifold models.
That should be reconfirmed by concrete calculations showing that all Yukawa
couplings \cite{Yukawa,KO,Yukawa2} have no complex phase except non-physical
phases.
We discuss the $Z_6$ orbifold model, whose nontrivial states
(5) might lead to complex Yukawa couplings.
For example we consider couplings of the $\theta^3$-twisted state (5) with
$\theta$-twisted and $\theta^2$-twisted sectors.
We take $|(\theta^2,0)>$ as the $\theta^2$-twisted state, while the
$\theta$-twisted sector has only the state $|(\theta,0)>$ as the
ground state.
The selection rules of the Yukawa couplings of these states are investigated
in ref.\cite{KO}.
Each state associated with the fixed point in (5) can couple to
$|(\theta,0)>$ and $|(\theta^2,0)>$.
In addition these derive the same coupling strength.
Therefore the couplings of these state has a factor concerning with phase as
$$(1+\exp [2i\gamma ]+\exp [i \gamma ]),
\eqno(12)$$
up to overall phases to define these states.
This coupling vanish unless $ \exp [i \gamma]=1$.
As a result this coupling has no physical complex phase.
In a similar way we can show that Yukawa couplings of the orbifold models do
not have any physical complex phase \cite{Yukawa2}.

\vspace{0.8 cm}
\leftline{\large \bf 4. Conclusion}
\vspace{0.8 cm}

We have studied the CP symmetry in the superstring theories from
geometrical viewpoint.
It turns out that C and P are 16-dimensional and 10-dimensional orientation
preserving operators, respectively.
We have shown the well defined P operators always exist in the orbifold
models as Weyl reflections, which are consistent with the Wilson lines and
complicated structures of the 6-dimensional space like (5).
Although here we have restricted ourselves to the 6-dimensional orbifolds
which are the products of the 2-dimensional orbifolds, similarly we can study
CP of the other $Z_N$ orbifold models as well as $Z_N \times Z_M$ orbifold
models.
However we have to take into account the fact that for the $Z_N \times Z_M$
orbifold models there is another contribution to $\Delta$ and it is called
discrete torsion \cite{Vafa,ZMN1}.

Our result of no CP violating
complex Yukawa coupling in orbifold models demands spontaneous CP violation
below a compactification scale.
For example some complex phases may appear when the supersymmetry is broken.
In ref.\cite{Choi} it is shown that symmetries of dilaton and moduli fields
in an effective lagrangian suppress the appearance of the CP phases at the
supersymmetry breaking.
Thus it is important to study further how the symmetries of orbifold models
including CP discussed in the present paper might affect the CP violation
in the effective lagrangians.

\noindent {\large{\bf Acknowledgment}}

The authors would like to thank M.~Konmura, M.~Sakamoto, D.~Suematsu,
K.~Yamada and Y.~Yamagishi for their very valuable discussions.
The work of T.K. is partially supported by Soryuushi Shogakukai.

\end{document}